\newcommand{\PRL}[3]{Phys.\ Rev.\ Lett.\ {\bf #1},\ #2 (#3)}

\newcommand{\NAT}[3]{Nature (London)\ {\bf #1},\ #2 (#3)}

\newcommand{\PRA}[3]{Phys.\ Rev.\ A\ {\bf #1},\ #2 (#3)}

\renewcommand\a{\alpha}

\newcommand{\ran}{\rangle}

\newcommand{\diracslash}[1]{#1\llap{/\kern2pt}}

\newcommand{\be}{\begin{equation}}
\newcommand{\ee}{\end{equation}}
\newcommand{\bea}{\begin{eqnarray}}
\newcommand{\eea}{\end{eqnarray}}
\newcommand{\ba}[1]{\begin{array}{#1}}
\newcommand{\ea}{\end{array}}

\documentclass[floats,secnumarabic,nobibnotes,nofootinbib,aps,prl,showpacs]{revtex4}
\usepackage{graphicx}
\usepackage{color}
\begin{document}

\preprint{APS/123-QED}

\title{Entanglement induced Sub-Planck structures }

\author{Jitesh R. Bhatt}
\email{jeet@prl.res.in}
\author{Prasanta K. Panigrahi}
\email{ prasanta@prl.res.in}
\author{Manan Vyas}
\email{manan@prl.res.in}

 \affiliation{ Physical Research Laboratory, Navrangpura, Ahmedabad 380 009,
India}        

\date{\today}% It is always \today, today,
\begin{abstract}

We study Wigner function of a system describing entanglement of two
superposed coherent-states.
Quantum interferece arising due to entanglement is shown to produce
sub-Planck structures in the phase-space plots of the Wigner function. 
Origin of these structures in our case depends on entanglement 
unlike those in Zurek \cite{Zurek}. 
It is argued that these kind of entangled compass states
are better suited for carrying out precision measurements.
\end{abstract}
% PACS, the Physics and Astronomy
                             % Classifiion Scheme.

\maketitle
%\section{Introduction}
Recently it has been demonstrated that, the quantum states obtained from superposition of
coherent states can be useful in quantum metrology, especially in carrying 
out Heisenberg-limited measurements
and quantum parameter estimation [1,2]. Zurek has shown that non-local superposition
of coherent quantum states can have well-defined oscillatory structures in the 
phase space, at scales smaller than Planck constant
$\hbar$. More surprisingly, contrary to the
commonly held belief, these structures can be physically important. 
The Wigner distribution function  for a compass state,
 $|\alpha\rangle+|-\alpha \rangle+|i\alpha \rangle+|-i\alpha \rangle$
has checker-board type of structure in the phase space due to quantum
interference. Here $\a$ is a complex parameter used for characterizing the 
coherent states and its magnitude, 
in the present context, signifies a distance from the origin in the phase-space.
The typical area $a$ of the fundamental tile of the checker-board is
$a=\hbar\frac{\hbar}{\cal A}$ where, ${\cal A}$ is the area of the accessible phase
space, which can be estimated from the total energy. The sub-Planck structures arise
because for a compass state with  superposition of well-separated coherent states in the 
phase-space,  $\frac{\hbar}{\cal A}\ll 1$ and thus  $a\ll \hbar$. 
Locations of the coherent states in the phase-space of the compass state can be
denoted by, intuitively obvious, geographical notations
namely north ($N$), south ($S$), east ($E$) 
and west ($W$), to denote their relative positions in the phase space.
It can be shown that the interference between $NS(|i\alpha \rangle+|-i\alpha \rangle)$ and 
$EW(|-\alpha\rangle+|\alpha \rangle)$ combinations
 produce the checker-board type of pattern \cite{Zurek}.
If an act of measurement or any other process  displaces the original compass state by a
distance $\sqrt{a}$ in the phase space, the sub-Planck structures allow
the resultant state to be distinguished from the old one. i.e., both the states 
become  approximately orthogonal.
Thus such states are argued to be sensitive to external perturbations.
Quite naturally they can have implications for processes like decoherence 
 \cite{Zurek}. These states are useful in carrying out Heisenberg-limited sensitive
measurements \cite{Toscano}. 
It ought to be noted here that a simpler state involving just two superposed coherent states
can also  be used for carrying out these measurements. However such states offer sensitivity
to the perturbations in only one direction i.e., in the direction perpendicular to 
the line joining the two coherent states \cite{Toscano}. 
The Wigner function for these kind of
states do not show the checker-board patteren in phase space. 
In comparision the compass state, can offer sensitivity in all directions of the phase-space.

 The sub-Planck structures have been studied by various researchers
to further investigate their properties and test some of the assumptions made. 
The issue of sensitivity to external perturbation of these structures, as the system evolves
with time, was studied using Loschmidt echoes \cite{Beena}. However the results of this work remain inconclusive. Recently it has been demonstrated that the above compass states, due to their
Heisenberg-limited sensitivity to external perturbation can be utilized for quantum parameter
estimation \cite{Toscano}. 
In this work, the state was entangled to a two-level atomic system for carrying out
the measurements.
Generation of the compass state in cavity QED and its decoherence characteristics 
were studied in Ref.\cite{GSA}. Such states can also be generated during the fractional revival
process of molecular wave-packets \cite{Ghosh}. 
A classical analog of these states was found in Ref.\cite{Praxmeyer}. It was shown that two pulses displaced by a small sub-Fourier shift of the carrier frequency become mutually orthogonal.
The single particle compass state used in Ref.\cite{Zurek} is composed of
four coherently superposed localized Gaussians. These states 
are rather difficult to produce. It has been shown in Ref.\cite{GSA} that, the
interference pattern (sub-Planckian structures) arising due to the
superposition of all the four Gaussians ($NS$ and $EW$ combinations) can
disappear faster than the interference pattern between any two Gaussians
due to decoherence.
In this work, we  aim to study the sub-Planck structures in the phase space of
a bipartite system and analyse the role of entanglement in it.
For this goal, we  propose a new compass
state $\vert\psi\rangle_c$ consisting of the following:
\begin{equation}
\vert\psi\ran_c=\frac{1}{\sqrt{2}}(A\vert \pm\a \ran_1\vert \pm\iota\a \ran_2
+ B\vert \pm\iota\a \ran_1\vert \pm\a \ran_2),
\end{equation}
where, $A=A_1+iA_2$ and $B=B_1+iB_2$ are complex parameters that control the entanglement. The states in Eq.(1) are given by,

\be
\vert \pm\a \ran=\frac{1}{\sqrt{2}}(\vert \a\ran\,+\,\vert-\a\ran).
\ee

The choice of the compass state $\vert\psi\rangle_c$  is such that, when 
one considers  the Wigner function for a constituent particle state (i.e., for state like in Eq.(1)), it 
does not show any checker-board pattern.
But the Wigner function for the entire $\vert\psi\rangle_c$ shows these structures. 
It can be demonstrated that they  arise solely due to entanglement. 
Since the degree of entanglement in the state given by Eq.(1) is determined by $A$
and $B$ only and does not depend on $\a$, there are two different  
decoherence characteristics for the superposition and the entanglement. 
Superpostion of the coherent state is strongly affected by
noise process say absorption of photons, while it can have no effect on the entanglement
\cite{Enk}. The entangled coherent states are more
robust against the decoherence arising due to photon absorbtions noise.
Thus keeping in mind the results of Refs.\cite{GSA, Enk}, the state proposed by us can be 
more suitable for carrying out Heisenberg-limited measurements.
Since the proposed entangled states give sub-Planck structures in the Wigner function, they
offer sensitivity in all the directions in the phase-space.

We represent the states states (given in Eq.(1))  by
localized coherent Gaussian states, to construct the normalized coordinate $\vert \pm\a\ran\rightarrow\psi(x)$ 
and the momentum even states $\vert \pm i\a\ran\rightarrow\varphi(x)$:
\begin{equation}
\psi(x)=\frac{e^{-(x+x_{0})^{2}/2\delta^{2}}+e^{-(x-x_{0})^{2}/2\delta^{2}}}
{\sqrt{2}\pi^{1/4}\delta^{1/2}\left[ 1+e^{-x_{0}^{2}/\delta^{2}}\right]^{1/2}}
\end{equation}
and,
\begin{equation}
\varphi(x)=\frac{e^{-x^{2}/2 \delta^{2}+\iota p_{0} x/
\hbar}+e^{-x^{2}/2 \delta^{2}-\iota p_{0} x/
\hbar}}{\sqrt{2}\pi^{1/4}\delta^{1/2}\left[
1+e^{-p_{0}^{2}\delta^{2}/\hbar^{2}}\right]^{1/2}},
\end{equation}
\noindent
where, $x_{0}$, $p_{0}$ and $\delta$ are taken to be real quantities.
Superpostion of the states given by Eqs.(3-4) can give the representation of the  single particle compass state considered in Zurek \cite{Zurek}. 
The compass state proposed here is given by,
\begin{equation}
\Psi(x_{1},x_{2})={\cal N}[A \psi(x_{1}) \phi(x_{2}) + B \phi(x_{1})\psi(x_{2})]
\end{equation}
where, ${\cal N}$ is the normalization constant.
It should be mentioned that 
non-separability condition for the wavefunctions of continuous variables is not fully established. 
The state in Eq.(5) does not satisfy separability criterion  based
on the variance approach \cite{Simon,Duan}.

From Eq.(5) the correlation function is obtained,
\be
 c(x_1,a_1,x_2,a_2)=\Psi^{\dagger}\left( x_{1}+\frac{a}{2},x_{2}+\frac{b}{2}\right)\\
\Psi\left( x_{1}-\frac{a}{2},x_{2}-\frac{b}{2}\right).
\ee
\noindent

The Wigner function, in four dimensional phase-space, can then be defined as,
\begin{equation}
W(x_{1}, p_{1} ; x_{2}, p_{2})
=\frac{1}{(2\pi\hbar)^{2}}{\int_{-\infty}^{\infty}}{\int_{-\infty}^{\infty}}c(x_1,a_1,x_2,a_2)
e^{\frac{i(p_{1}a+p_{2}b)}{\hbar}}da db  .
\end{equation}
A lengthy calculation yeilds

\bea
W(x_{1}, p_{1} ; x_{2}, p_{2})&=&
\frac{2\delta^{2}c|{\cal N}|^{2}}{\pi\hbar^{2}}e^{-\frac{(x_{1}^{2}+x_{2}^{2})}{\delta^{2}}-\frac{(p_{1}^{2}+p_{2}^{2})\delta^{2}}{\hbar^{2}}}
(W_{D1}+W_{D2}+ \nonumber \\ & &e^{-\frac{x_{0}^{2}}{2\delta^{2}}-\frac{p_{0}^{2}\delta^{2}}{2\hbar^{2}}}(W_{C1}+W_{C2})),
\eea
where, $W_{D1}$, $W_{D2}$ and  $W_{C1}$, $W_{C2}$ are, respectively, the
diagonal and off-diagonal components of the Wigner function.

 First consider one of the diagonal terms,
\begin{widetext}
\bea
W_{D1}&=&2|A|^{2}(e^{-\frac{x_{0}^{2}}{\delta^{2}}-\frac{p_{0}^{2}\delta^{2}}{\hbar^{2}}}\cosh\left(\frac{2p_{0}p_{2}\delta^{2}}{\hbar^{2}}
\right) \cosh\left(\frac{2x_{0} x_{1}}{\delta^{2}} \right)+
e^{-\frac{x_{0}^{2}}{\delta^{2}}}\cosh\left(\frac{2x_{0}
x_{1}}{\delta^{2}} \right)\cos\left(
\frac{2p_{0}x_{2}}{\hbar}\right)+\nonumber \\ & &
e^{-\frac{p_{0}^{2}\delta^{2}}{\hbar^{2}}}\cos\left( \frac{2x_{0}
p_{1}}{\hbar}\right)\cosh\left(\frac{2p_{0}p_{2}\delta^{2}}{\hbar^{2}}
\right) +2\cos\left( \frac{2p_{0}x_{2}}{\hbar}\right)\cos\left(
\frac{2x_{0} p_{1}}{\hbar}\right)).
\eea
\end{widetext}
It can be seen from above that the first three terms containing hyperbolic functions 
are multiplied by constant Gaussian factors which are bound to be small, in
the present mesoscopic context concerned with relatively larger values of $x_0$ and $p_0$.
Thus only the last term in Eq.(9) becomes dominant in the region between the Gaussians. 
This term is a purely oscillating term, which can produce  significant amount of
interference. The zeroes of this term 
occur at $x_{2}=\pm\frac{\pi\hbar}{4p_{0}}$ and
$p_{1}=\pm\frac{\pi\hbar}{4x_{0}}$ from which one can calculate the fundamental
area of the  tile
as  $\frac{(2\pi\hbar)^{2}}{4x_{0} p_{0}}$. 
It should be noted that Eq.(9) has a
$|A|^2$ factor and the above calculation does not require any information about entanglement.
Indeed one can see sub-Planck structure 
in this plane
even when $|B|=0$. As this plane has mixed coordinates i.e., momentum $p_1$ of particle one and position $x_2$ of
particle two,  we believe that the structures observed here are not physically 
important.

Now we  compute the other diagonal term which has $|B|^2$ as a factor,
\begin{widetext}
\bea
W_{D2}&=&2|B|^{2}(
e^{-\frac{x_{0}^{2}}{\delta^{2}}-\frac{p_{0}^{2}\delta^{2}}{\hbar^{2}}}\cosh\left(\frac{2p_{0}p_{1}\delta^{2}}{\hbar^{2}}
\right)  \cosh\left(\frac{2x_{0} x_{2}}{\delta^{2}}
\right)+e^{-\frac{x_{0}^{2}}{\delta^{2}}}\cosh\left(\frac{2x_{0}
x_{2}}{\delta^{2}} \right)\cos\left(
\frac{2p_{0}x_{1}}{\hbar}\right)+\nonumber \\ & &
e^{-\frac{p_{0}^{2}\delta^{2}}{\hbar^{2}}}\cos\left( \frac{2x_{0}
p_{2}}{\hbar}\right)\cosh\left(\frac{2p_{0}p_{1}\delta^{2}}{\hbar^{2}}
\right) +2\cos\left( \frac{2p_{0}x_{1}}{\hbar}\right)\cos\left(
\frac{2x_{0} p_{2}}{\hbar}\right)).
\eea
\end{widetext}

This has a similar structure as that of $W_{D1}$ except that it has  two
Gaussians located at $x_{2}=x_{0}$ and 
$p_{1}=p_{0}$. From the argument given above one can see the sub-Planck structures
in the $x_1p_2$-plane. Once again they may not be physically relevant.

Next we examine the off-diagonal terms, which can be computed to be
 
\begin{widetext}
\bea
W_{C1}&=&((A_{1}B_{1}+A_{2}B_{2})-(A_{1}B_{2}-A_{2}B_{1}))( e^{\frac{i
p_{0}x_{0}}{\hbar}}(\cosh\left( (\frac{x_{0}}{\delta^{2}}-\frac{i
p_{0}}{\hbar})(x_{1}+x_{2})+(\frac{ix_{0}}{\hbar}-\frac{p_{0}\delta^{2}}{\hbar^{2}})(p_{1}-p_{2})\right)
+\nonumber \\ & & \cosh\left( (\frac{x_{0}}{\delta^{2}}-\frac{i
p_{0}}{\hbar})(x_{1}-x_{2})+(\frac{ix_{0}}{\hbar}-\frac{p_{0}\delta^{2}}{\hbar^{2}})(p_{1}+p_{2})\right))+\nonumber
\\ & & e^{-\frac{i p_{0}x_{0}}{\hbar}}(\cosh\left(
(\frac{x_{0}}{\delta^{2}}+\frac{i
p_{0}}{\hbar})(x_{1}+x_{2})+(\frac{ix_{0}}{\hbar}+\frac{p_{0}\delta^{2}}{\hbar^{2}})(p_{1}-p_{2})\right)+\nonumber
\\ & & \cosh\left( (\frac{x_{0}}{\delta^{2}}+\frac{i
p_{0}}{\hbar})(x_{1}-x_{2})+(\frac{ix_{0}}{\hbar}+\frac{p_{0}\delta^{2}}{\hbar^{2}})(p_{1}+p_{2})\right)
)+\nonumber \\ & &
2(\cos\left(p_{0}(\frac{(x_{1}-x_{2})}{\hbar}-\frac{i
(p_{1}+p_{2})\delta^{2}}{\hbar^{2}}) \right)\cosh\left(
x_{0}(\frac{(x_{1}+x_{2})}{\delta^{2}}+\frac{i(p_{1}-p_{2})}{\hbar}
)\right)+\nonumber \\ & &
\cos\left(p_{0}(\frac{(x_{1}+x_{2})}{\hbar}-\frac{i
(p_{1}-p_{2})\delta^{2}}{\hbar^{2}}) \right)\cosh\left(
x_{0}(\frac{(x_{1}-x_{2})}{\delta^{2}}+\frac{i(p_{1}+p_{2})}{\hbar}
)\right)))
\eea
\end{widetext}

and,

\begin{widetext}
\bea
W_{C2}&=&((A_{1}B_{1}+A_{2}B_{2})+(A_{1}B_{2}-A_{2}B_{1}))( e^{\frac{i
p_{0}x_{0}}{\hbar}}(\cosh\left((\frac{x_{0}}{\delta^{2}}-\frac{i
p_{0}}{\hbar})(x_{1}+x_{2})-(\frac{ix_{0}}{\hbar}-\frac{p_{0}\delta^{2}}{\hbar^{2}})(p_{1}-p_{2})\right)+\nonumber
\\ & & \cosh\left( (\frac{x_{0}}{\delta^{2}}-\frac{i
p_{0}}{\hbar})(x_{1}-x_{2})-(\frac{ix_{0}}{\hbar}
-\frac{p_{0}\delta^{2}}{\hbar^{2}})(p_{1}+p_{2})\right) )+\nonumber \\ & &
e^{-\frac{ip_{0}x_{0}}{\hbar}}(\cosh\left(
(\frac{x_{0}}{\delta^{2}}+\frac{i
p_{0}}{\hbar})(x_{1}+x_{2})-(\frac{ix_{0}}{\hbar}+\frac{p_{0}\delta^{2}}{\hbar^{2}})(p_{1}-p_{2})\right)+\nonumber
\\ & & \cosh\left( (\frac{x_{0}}{\delta^{2}}+\frac{i
p_{0}}{\hbar})(x_{1}-x_{2})-(\frac{ix_{0}}{\hbar}+\frac{p_{0}\delta^{2}}{\hbar^{2}})
(p_{1}+p_{2})\right) )+\nonumber \\ &
&2(\cos\left(p_{0}(\frac{(x_{1}-x_{2})}{\hbar}+\frac{i
(p_{1}+p_{2})\delta^{2}}{\hbar^{2}}) \right)\cosh\left(
x_{0}(\frac{(x_{1}+x_{2})}{\delta^{2}}-\frac{i(p_{1}-p_{2})}{\hbar}
)\right)+\nonumber \\ &
&\cos\left(p_{0}(\frac{(x_{1}+x_{2})}{\hbar}+\frac{i
(p_{1}-p_{2})\delta^{2}}{\hbar^{2}}) \right)\cosh\left(
x_{0}(\frac{(x_{1}-x_{2})}{\delta^{2}}-\frac{i(p_{1}+p_{2})}{\hbar}
)\right) ))).
\eea
\end{widetext}

Interestingly one sees the presence of EPR variables in each term. 
However, one finds that although purely oscillatory terms are present here,
they are significantly damped  as compared to the diagonal terms for large values
of $x_0$ and $p_0$. This is clearly evident from the Wigner function.

From the discussion so far, one may wonder if it is possible at all to  see checker-board type sub-Planck structures in $x_1p_1$ or $x_2p_2$ planes. 
Answer to this question can be found by adding the oscillatory terms from Eqs.(9-10),
\be
  4|A|^2\cos\left( \frac{2p_{0}x_{2}}{\hbar}\right)\cos\left(
\frac{2x_{0} p_{1}}{\hbar}\right) + 4|B|^2\cos\left( \frac{2p_{0}x_{1}}{\hbar}\right)\cos\left(
      \frac{2x_{0} p_{2}}{\hbar}\right).
\ee
From the above, distance between two zeros in $x_1$ direction is again $\pm\frac{\pi\hbar}{4p_{0}}$ 
while it is  $\pm\frac{\pi\hbar}{4x_{0}}$ in $p_1$-direction. This gives the area of the fundamental
tile $a=\frac{(2\pi\hbar)^{2}}{4x_{0} p_{0}}$ in $x_1p_1$ plane of particle one. Similarly one can find zeros in $x_2$
and $p_2$ directions and obtain the same value of the fundamental area. It should be noted that fundamental area $a$, though
does not depend upon $A$ or $B$,  both of them need to be  simultaneously non zero in
order to get sub-Planck structures in the physical  $x_1p_1$ or $x_2p_2$ plane. 
It is clear that visibility of the interference patterns depends upon the relative magnitudes
of $A$ and $B$. 
%-----------------------Fig.(1) In Tabular Form-------------
\begin{figure*}
\begin{tabular}{cc}
\scalebox{0.21}{\includegraphics{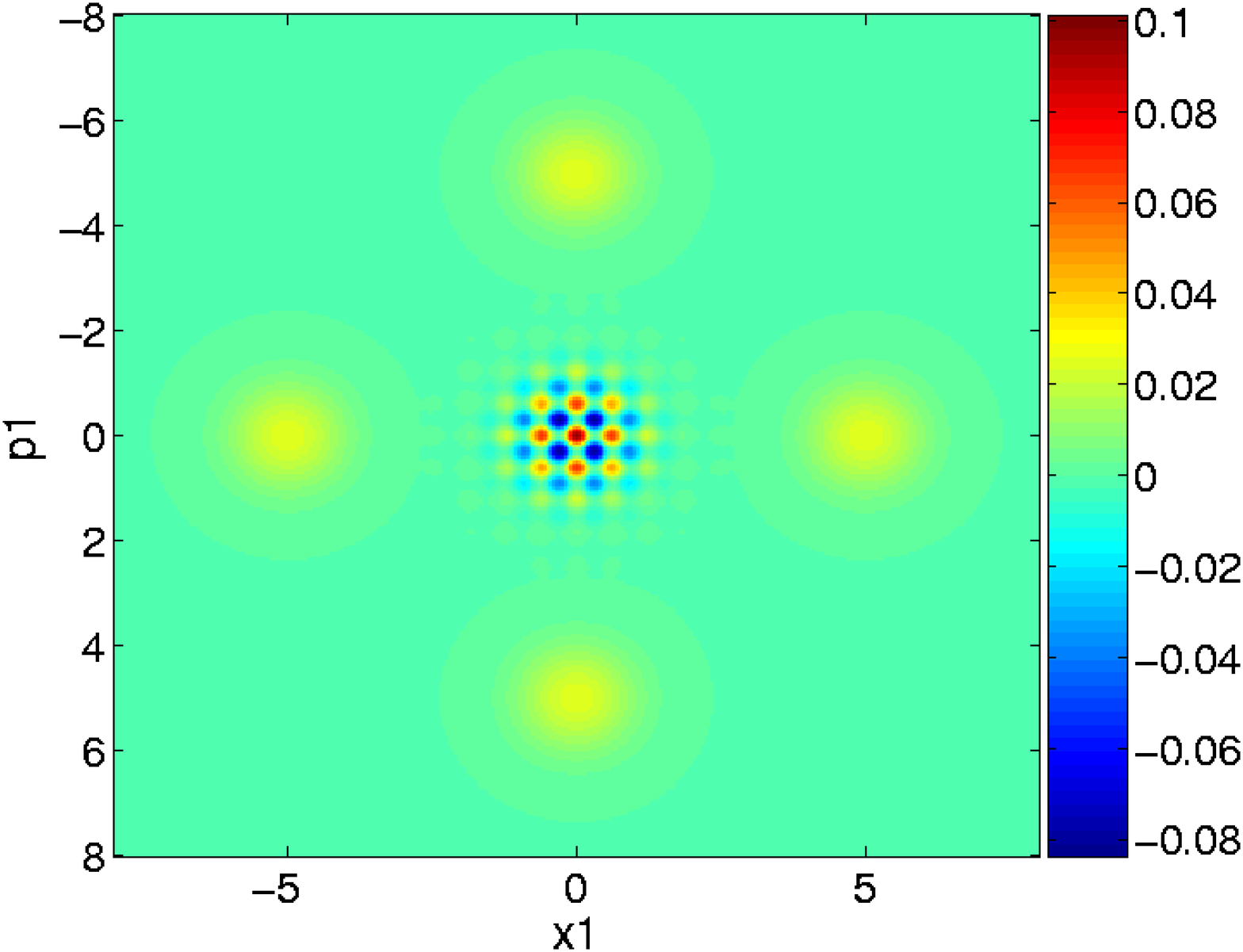}}\\
\scalebox{0.21}{\includegraphics{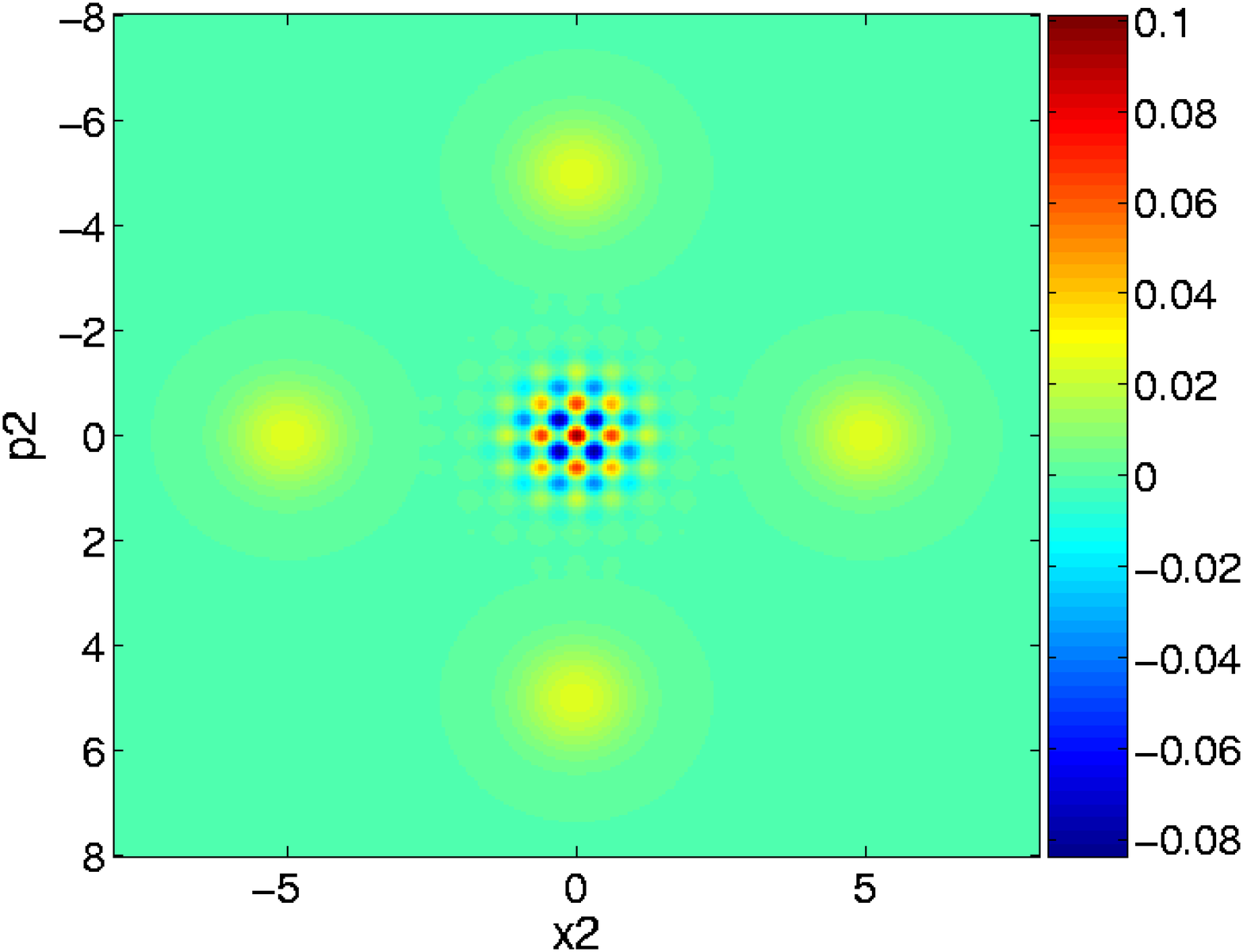}}&
\end{tabular}
\caption{\label{SPT}(color online) Cross-sectional view of Wigner function of entangled
wavefunction with $A_1=A_2=\frac{1}{\sqrt{2}}$ and 
$B_1=-B_2=\frac{1}{\sqrt{2} }$  in (a) $(x_{1} p_{1})$ plane, with
$x_2=0,\, p_2=0$ and (b)
$(x_{2}p_{2})$ plane with $x_1=0,\, p_1=0$.}
\end{figure*}
%-----------------------------------------------------
%-----------------------Fig.(2) In Tabular Form-------------
\begin{figure*}
\begin{tabular}{cc}
\scalebox{0.21}{\includegraphics{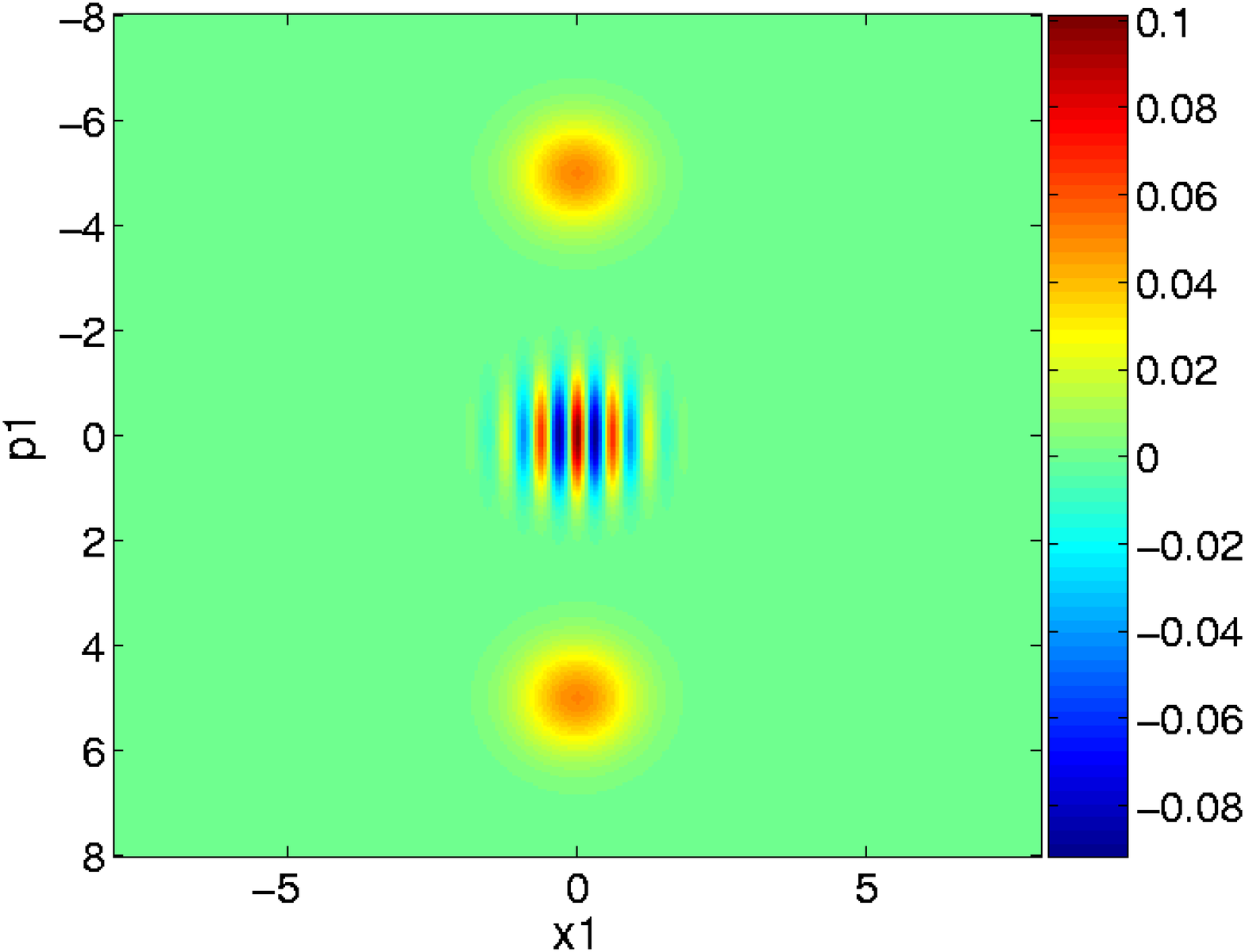}}\\
\scalebox{0.21}{\includegraphics{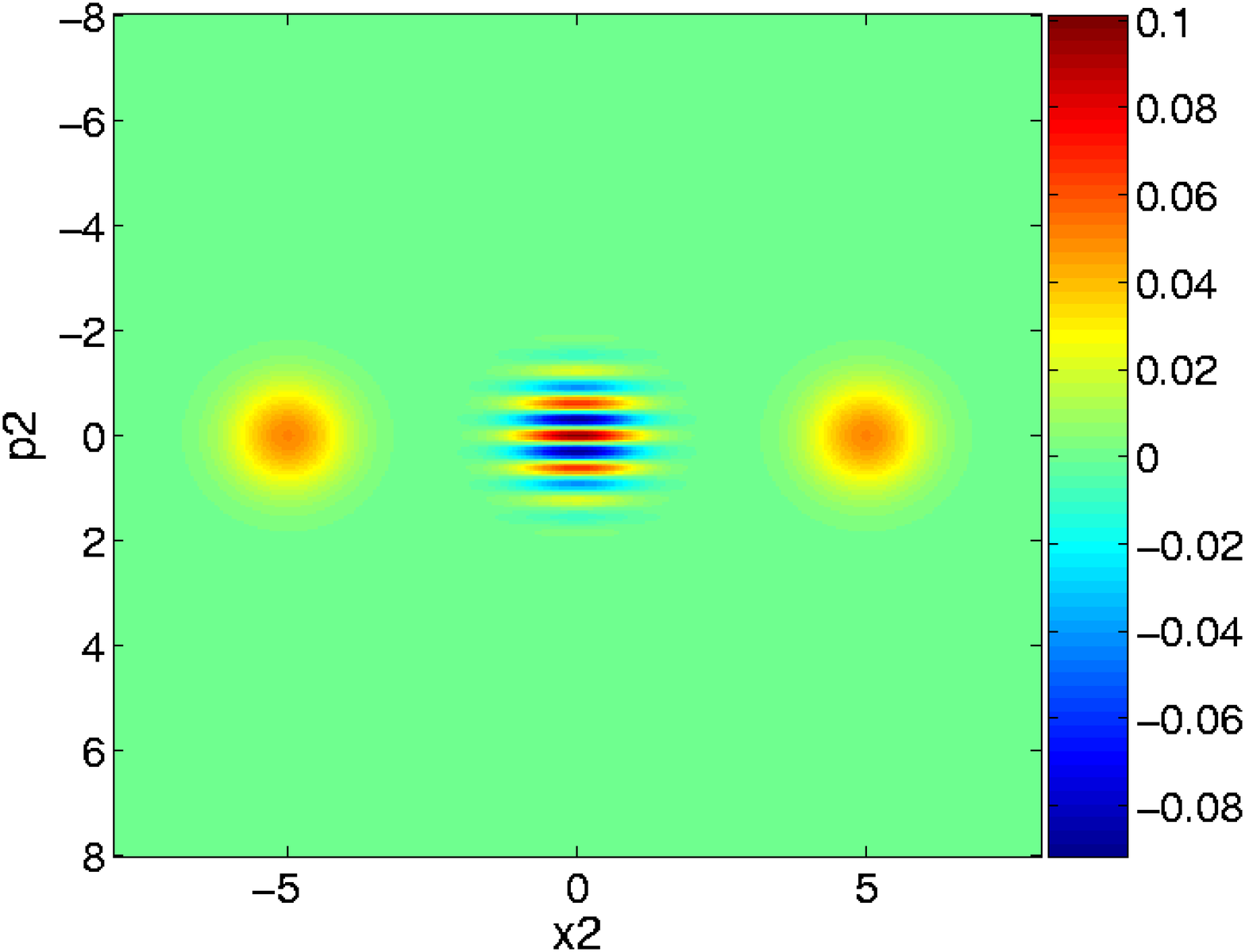}}&
\end{tabular}
\caption{\label{SPT}(color online) Cross-sectional view of Wigner function with parameter
value as in Fig.(1) except 
the entanglement has been turned off by setting
the parameter $B=0$.} 
\end{figure*}
%-------------------------------------------------
In  Figs.(1-2) we have shown plots of the 
Wigner function (Eq.(8))  in $x_1p_1$ and  $x_2p_2$ planes. Fig.(1)
depicts cross-sectional view of  the 
Wigner function
in $x_1p_1$ and $x_2p_2 $ planes while keeping  $A$ and $B$  both non
zero. It clearly shows the checker-board
type pattern with $a\ll\hbar$.  Area of the fundamental tile matches with
$a$ that we have calculated above. These 
plots look
very similar to that in \cite{Zurek}, but no oblique sidebands,  as
seen in  \cite{Zurek}, is visible in our figure. The oblique side bands
in our case come from the off diagonal terms $W_{C1}$ and $W_{C2}$. As
seen in Eq.(9) they are multiplied by constant Gaussian factors and
their contribution to the Wigner function is strongly suppressed
for sufficiently large values of $x_0$ and $p_0$. 
In this sense, we observe a cleaner checker-board type pattern
using the bipartite compass state.

Fig.(2) depicts the
case when $B=0$ and all the other parameters are same as in Fig.(1). 
No sub-Planck structure is seen here, 
which confirms our assertion that both $A$ and $B$ must be simultaneously non-zero to have these phase-space structures in $x_1p_1$ and $x_2p_2$ planes.
We have chosen $x_0$, $p_0$=5, in the units of $\hbar=1$ and
$\delta=1$ in plotting Figs.(1-2). The circles indicate the positions
of the Gaussians. For $B=0$ case, only two Gaussians are visible.

Sensitivity of the  entangled compass state in Eq.(3) can be studied
as follows. Let $D_1(\alpha)$ and  $D_2(\beta)$ denote two displacement
operators causing the displacent of particle states one and two, by amount $\alpha$
and $\beta$ respectively to create a perturbed state $\vert\psi_{per}\rangle=D_1(\alpha)D_2(\beta)
\vert\psi_c\rangle$.  The overlap function $\vert \langle\psi_c\vert\psi_{per}\rangle\vert^2$
can be found to be
%\begin{widetex}
\bea
\vert \langle\psi_c\vert\psi_{per}\rangle\vert^2&=&16\vert{\cal N}\vert^4\left[
\vert A\vert^4cos^2\left\{x_0(\beta+\beta^*)\right\}cosh^2\left\{x_0(\alpha^*-\alpha)\right\}
+\vert B\vert^4cos^2\left\{x_0(\beta+\beta^*)\right \}cosh^2\left\{x_0(\alpha^*-\alpha)\right\}
\right.
\nonumber \\
 \nonumber \\ & &
\left.
+2\vert A\vert \vert B\vert cos\left \{x_0(\beta+\beta^*)\right \}cosh\left\{x_0(\alpha^*-\alpha) \right \}
cos\left\{x_0(\beta+\beta^*)\right \}cosh\left\{x_0(\alpha^*-\alpha)\right\} 
\right].
\eea
%\end{widetex}
It is particularly of interest to
consider the case when there is equal shift to both the particles i.e., $\alpha=\beta=is\frac{x_0}{\vert x_0\vert} $, the overlap function can then be written as
\bea
8\vert{\cal N}\vert^4\left\{1+cos\left(4x_0s\right)\right\}. \nonumber
\eea
\noindent 
Clearly the overlap function becomes minimum, for the  distinguishable displacement,
if $s\sim\pi/(4x_0)$.
Next consider the displacement of the compass state given in Ref.\cite{Zurek,Toscano},
by amount $s_1$,
the overlap function can be written as
\bea
\frac{1}{4}\left\{3\,+\,4cos\left(2x_0s_1\right)\,+\,cos\left(4x_0s_1\right) \right\}
\eea
\noindent
This function becomes minimum for $s_1\sim\pi/(2x_0)$. Thus the distinguishable 
displacement coming from the entangled state is a factor $1/2$ less than the one 
found in Ref.\cite{Toscano}. Since the sensitivity to the measurement depends upon 
$\frac{1}{x_0}$, both the states, the one given in Refs.\cite{Zurek, Toscano} and
in Eq.(3), can be useful in carrying out the Heisenberg limited measurements.
The energy resource required by the
present state is less than  that required in the nonentangled scenario.

In conclusion, we have studied the phase-space structures in a bipartite system of
entangled superposed coherent-states. It was shown that the Wigner function for the
quantum state have sub-Planck structures arising due to entanglement. 
The Wigner function,  in the four dimensional phase-space, have these structures in
$x_1p_1$, $x_1p_2$, $x_2p_1$ and $x_2p_2$ planes. 
But we have argued that  patterns seen in the off-diagonal planes in the
phase-space may not be physically relevant. They exist with or without
entanglement. The structures seen in the diagonal planes are induced by entanglement and can be
physically important. Entanglement makes them a better compass state in the sense that
they are robust against decoherence. Furthermore, these structures are cleaner in this bipartite
system due to the suppression of the 
side bands. 
We have shown that this kind of compass state may be useful in carrying out 
precesion quantum measurements with less energy resource.


\newpage

\begin{references}

\bibitem{Zurek} W.H. Zurek, \NAT {412}{712}{2001}.
\bibitem{Toscano} F. Toscano, D.A.R. Dalvit, L. Davodovich and W.H. Zurek, \PRA {73}{023803}{2006}.


\bibitem{Beena} Ph. Jacquod, I. Adagideli and C.W.J Beenakkar, \PRL{89}{154103-3}{2001}.


\bibitem{GSA} G.S. Agarwal and P.K. Pathak, \PRA {70}{053813}{2004}.

\bibitem{Ghosh} S. Ghosh, A. Chiruvelli, J. Banerji and P.K. Panigrahi, \PRA {73}{013411}{2006}.

\bibitem{Praxmeyer} L. Praxmeyer, P. Wasylczyk, C. Radzewicz and
  K. W\`odkiewicz, \PRL{98}{063901}{2007}.

\bibitem{Enk} S.J. van Enk and O, Hirota, \PRA{64}{022313}{2001}.

\bibitem{Duan} L.-M. Duan, G. Giedke, J.I. Cirac and P. Zoller, \PRL {84}{2722}{2000},
E. Shchukin and W. Vogel, \PRL{95}{230502}{2005}.

\bibitem{Simon} R. Simon, \PRL {84}{2726}{2000}.


\end{references}
\end{document}